\documentclass[usenatbib,useAMS]{mn2e}
\usepackage{deluxetable}
\usepackage{epsfig}

\newcommand{\nuc}[2]{\ensuremath{\mathrm{^{#1}#2}}}

\newcommand{\msun}{\ensuremath{M_\odot}}
\newcommand{\simgt}{\,\hbox{\lower0.6ex\hbox{$\sim$}\llap{\raise0.6ex\hbox{$>$}}}\,}
\title[Effect of Auger and IC $e^-$ on SN light curves]{Late-time supernova light curves: The effect of internal conversion and Auger electrons}

\author[Seitenzahl et al. 2009]{I. R. Seitenzahl$^{1}$, S. Taubenberger$^{1}$, S. A. Sim$^{1}$\\
$^{1}$Max-Planck-Institute for Astrophysics,
                 85741 Garching, Germany\\} 
\date{\today}
\pubyear{}
\begin{document} 
\maketitle
           
\begin{abstract}
Energy release from radioactive decays contributes significantly to supernova light curves.
Previous works, which considered the energy deposited by $\gamma$-rays
and positrons produced by \nuc{56}{Ni}, \nuc{56}{Co},\nuc{57}{Ni},
\nuc{57}{Co}, \nuc{44}{Ti} and \nuc{44}{Sc},
have been quite successful in explaining the light curves of both core
collapse and thermonuclear supernovae. 
We point out that Auger and internal conversion electrons 
together with the associated X-ray cascade, 
constitute an additional heat source.
When a supernova is transparent to
$\gamma$-rays, these electrons can contribute
significantly to light curves for reasonable nucleosynthetic yields.
In particular, the electrons emitted in the decay of \nuc{57}{Co},
which are largely due to internal conversion from a fortuitously low-lying 3/2$^-$ 
state in the daughter \nuc{57}{Fe}, constitute an additional
significant energy deposition channel.
We show that when the heating by these electrons is accounted for,
a slow-down in the lightcurve of SN 1998bw is naturally obtained
for typical hypernova nucleosynthetic yields. 
Additionally, we show that for generic Type Ia supernova yields, the
Auger electrons emitted in the ground-state to ground-state electron
capture decay of \nuc{55}{Fe} exceed the energy released by the
\nuc{44}{Ti} decay chain for many years after the explosion. 

\end{abstract}

\begin{keywords}
{nuclear reactions, nucleosynthesis, abundances ---
  supernovae: general --- supernovae: individual (SN 1998bw) --- white dwarfs}
\end{keywords}

\section{Introduction}
\label{sec:intro}
One of the great successes of nuclear astrophysics was
demonstrating
that the light curves of both thermonuclear (Type Ia) and, at least for late times,
core collapse (Type Ib/c, Type II) supernovae are powered by the decay
of radionuclides synthesized in the explosion. 
Quickly following the dynamic explosion and nucleosynthesis phase a supernova enters a state of homologous
expansion. The temperature drops during the expansion so that
nuclear fusion ceases and radioactive decay is the only means of 
changing the composition.
\citet{pankey62}, \citet{truran67} and \citet{colgate69} first
proposed the now widely accepted view that the energy liberated in the
decay of radioactive \nuc{56}{Ni} followed by the decay of
\nuc{56}{Co} to stable \nuc{56}{Fe} is the most important source for re-heating the supernova ejecta to temperatures
high enough to shine brightly in the optical part of the electromagnetic spectrum.  
Initially, the bulk of the heating is induced by the energetic $\gamma$-rays
produced in the decays which thermalize and deposit their energy via
a number of processes, including Compton scattering and photoelectric absorption. 
Due to the expansion, however, the column density decreases with time as $t^{-2}$,
and the ejecta becomes more and more transparent to these high energy photons. 

\nuc{56}{Co} decays predominantly via electron capture, but 19\% of all decays proceed via positron emission. 
The fundamental importance of this positron channel was first pointed out by \citet{arnett79}.
Initially it was assumed that
the positrons deposit all of their kinetic energy before forming positronium and annihilating locally \citep{axelrod80}.
With that assumption, for a generic Type Ia supernovae (SN Ia) model, the ejecta
are sufficiently transparent to $\gamma$-rays after about $\sim200$
days that the \nuc{56}{Co} positrons
dominate the energy deposition rate
\citep[e.g.][]{woosley89,sollerman04}. 
Subsequently, the idea of complete and local energy deposition has been questioned.
\citet{chan93} suggested that as early as a few hundred days after the explosion
positrons may begin to escape, especially in the presence of radially combed magnetic
fields \citep[see also][]{milne99}. 
However, it has also been argued \citep[e.g.][]{swartz95}, that 
the fraction of positrons which escape cannot be very large since 
the observed light
curves could then not be explained without an additional source of heating.

SN~1987A has demonstrated that light curves can be constructed for 
nearby supernovae even several years after the
explosion. This has led to the consideration of 
less abundant but
longer lived radioactive species. 
The most important of these is \nuc{57}{Co}, which is expected to be produced in significant amounts as the decay product of the short lived \nuc{57}{Ni}. 
Due to its relatively longer half-life (271.79 days) and the higher
opacity of the ejecta to the emitted $\gamma$-rays, it may dominate
the bolometric light curve, especially of core collapse supernovae, at
late times \citep[e.g.][]{pinto88,woosley89}. 
It is widely argued, that the only other radionuclide which
noticeably contributes to the bolometric light curve is \nuc{44}{Ti}
\citep{woosley80,kumagai89}. \nuc{44}{Ti} has an even longer half-life
of 58.9 years \citep{ahmad06}, and the short lived daughter \nuc{44}{Sc} has a strong positron
channel, which means that the \nuc{44}{Ti} decay chain can be the dominant
energy source at very late times \citep{kumagai89,woosley89}.  

In section \ref{sec:decay}, we review the physics of radioactive
decay, emphasizing the roles of 
internal
conversion 
(IC)
and Auger electron production. 
We point out that, while \nuc{57}{Co} has no positron channel, 
due to the nuclear structure of the daughter \nuc{57}{Fe}, a
significant amount of energy is released in the form of IC and Auger electrons.
In section \ref{sec:lc}, we summarize the current observational status of late
time supernova bolometric light curves of SNe Ia and SN~1998bw, a well observed
member of the broad-lined SNe~Ic class.
For both, we sketch the effect electrons emitted in radioactive decays
may have.  
Implications of our study are discussed and our conclusions are drawn in 
Section \ref{sec:discussion}. 

\section{Radioactive decay}
\label{sec:decay}

Most of the mass synthesized by fusion processes in supernova explosions consists of nuclear species in the mass range $A\approx 12-70$.
The nuclei are typically either stable or lie on the proton-rich side of stability. 
For the relevant nuclei, fission and $\alpha$-decay plays no role. 
The radioactive decay considered here occurs along an isobar towards more neutron rich nuclides in one of two ways, electron capture or positron emission. 
The first proceeds via the capture of an inner atomic (typically K or L shell) electron by a nuclear proton and the emission of an electron neutrino. 
The latter proceeds by the decay of a nuclear proton into a neutron and the emission of a positron and an electron neutrino. 
For both positron emission and electron capture, the transition to the daughter is a statistical process to a distribution of (excited) nuclear levels, with 
branching ratios given by the transition probabilities. 

Following electron capture, the daughter will have an electron hole in its atomic structure. 
Electrons cascade down to fill the gaps in the lower
lying atomic shells and typically a series of characteristic
fluorescence X-rays is emitted in the process. 
However, at every atomic transition, there is the possibility
that instead of emission of a photon, the excess energy is carried
away by one or several ejected outer electrons (Auger electrons). 
For ground-state to ground-state electron capture transitions, such as in the decay of \nuc{55}{Fe}, no $\gamma$-rays or positrons are emitted. 
Assuming that the neutrino escapes without interactions, in such transitions the Auger electrons and the X-rays constitute the only sources of radioactive heating.

 An excited state of the daughter nucleus typically transitions to lower lying states by emitting $\gamma$-ray photons until the ground state is reached.
For each transition, however, there exists the additional possibility to de-excite via IC. 
In this process, the energy difference of the nuclear levels is carried away by the ejection of an inner atomic electron and there is no $\gamma$-ray photon emitted. 
These electrons are called IC electrons. 
The probability to de-excite via IC, as measured by the IC coefficient 
$\alpha = \frac{\# \mathrm{of} \; e^- \; \mathrm{de-excitations}}{\# \mathrm{of} \; \gamma \; \mathrm{de-excitations}}$, is generally small, but increases for low lying states. 
In the decay of \nuc{57}{Co} in particular, IC
electrons are relatively copiously produced, due to the existence of a
low-lying nuclear level in the daughter \nuc{57}{Fe}.
For this 14.4 keV $3/2^{-}$ state of \nuc{57}{Fe}, IC
is in fact the preferred mode of de-excitation, with $\alpha = 8.58$. 

\section{Late-time bolometric light curves}
\label{sec:lc}
Late-time bolometric light curves have already been reconstructed from
multi-band photometry for several objects \citep[e.g.][]{suntzeff02,sollerman02b,sollerman04}.
For reliable reconstruction, the contribution of the UV/optical ($UBV\!RI$) and near-infrared ($JHK$) 
bands should be included (UVOIR light curve). However, given that near-IR observations are rare at very 
late epochs, sometimes only $B$-through-$I$ band observations are used, applying a
near-IR correction extrapolated from earlier epochs
\citep[e.g.][]{sollerman02b}.

Apart from the heating by various radioactive nuclei, a number of additional 
effects impact the slope of late-time SN light curves. Most importantly the ejecta become 
increasingly transparent to $\gamma$-rays \citep{leibundgut92} and -- much later -- to positrons 
\citep{ruizlapuente98,milne99}, as already mentioned above. This leads to a decline of the light 
curve steeper than expected for a scenario with a constant energy-deposition fraction. 
\citet{axelrod80} was the first to describe the `infrared catastrophe', which occurs 
when the temperature in the ejecta drops below a critical value. Thereafter, the 
cooling is mostly accomplished by fine-structure lines in the mid and
far infrared. While this 
is merely a re-distribution of the emission and has no effect on the true bolometric 
luminosity, it would show up as a faster decline of the UVOIR light curves 
constructed from observations with a limited wavelength range. Dust formation within 
the ejecta -- though a different physical process -- acts in a similar manner.
Conversely, a {\it slow-down} of the bolometric decline rate is observed if `freeze-out' 
occurs \citep{fransson93}, i.e., if additional heating is provided by the delayed 
recombination of electrons at late phases. Interaction of the ejecta with 
circumstellar material, resulting in a transformation of kinetic energy into light, 
can have a similar effect. The same is true for possible light 
echoes, where SN light from an earlier epoch is scattered off interstellar dust 
clouds towards the observer \citep{schmidt94,sparks99,patat06}.

To check for the presence or absence of these effects, it is important to model 
the contribution of the various radioactive decays to the bolometric luminosity 
evolution as accurately as possible, including the hitherto neglected IC and Auger electrons whose contributions
become important at very late epochs.
To illustrate the effects of these electrons, we consider the following four decay chains. 
\begin{eqnarray}
& &^{56}\mathrm{Ni}  \;\stackrel{t_{1/2} = \; 6.08d}{\hbox to 60pt{\rightarrowfill}} \; ^{56}\mathrm{Co} \; 
\stackrel{t_{1/2} = \; 77.2d}{\hbox to 60pt{\rightarrowfill}} \; ^{56}\mathrm{Fe} \\*
& &^{57}\mathrm{Ni}  \;\stackrel{t_{1/2} = \; 35.60 h}{\hbox to 60pt{\rightarrowfill}}\; ^{57}\mathrm{Co} \;
\stackrel{t_{1/2} = \; 271.79d}{\hbox to 60pt{\rightarrowfill}} \; ^{57}\mathrm{Fe}\\*
& &^{55}\mathrm{Co}  \;\stackrel{t_{1/2} = \; 17.53 h}{\hbox to 60pt{\rightarrowfill}}\; ^{55}\mathrm{Fe} \;
\stackrel{t_{1/2} = \; 999.67 d}{\hbox to 60pt{\rightarrowfill}} \; ^{55}\mathrm{Mn}\\*
& &^{44}\mathrm{Ti}  \;\stackrel{t_{1/2} = \; 58.9 y}{\hbox to 60pt{\rightarrowfill}}\; ^{44}\mathrm{Sc} \;
\stackrel{t_{1/2} = \; 3.97 h}{\hbox to 60pt{\rightarrowfill}} \; ^{44}\mathrm{Ca}
\end{eqnarray}

To circumvent the more complicated treatment of the photon transport,
in the subsequent discussion pertinent to very late-time bolometric light curves
of different classes of supernovae, 
we limit ourselves to comparing leptonic (and X-ray) energy injection
rates.  
We do not include the energy produced by the pair annihilation and
further assume that the kinetic energy of the leptons is completely
thermalized in situ. Thus, possible escape of positrons (or fast
electrons) is neglected.  

The energy generation rates presented here do not include any heating due to
$\gamma$-rays and therefore do not accurately predict bolometric light curves. 
Our approach however does allow for a direct comparison of the relative contributions to the heating of the
positrons produced in the decays of \nuc{56}{Co} and \nuc{44}{Sc} and
the electrons produced in the decays of \nuc{57}{Co} and \nuc{55}{Fe}. 
The relevant energies of the different decay channels are listed
in Table~\ref{tab:2}.
These data were extracted from the Chart of Nuclides database,  National Nuclear Data Center\footnote{http://www.nndc.bnl.gov/chart/}.

\begin{table}
\caption{Radioactive decay energies \label{tab:2} (keV decay$^{-1}$)}
\begin{tabular}{ccccc} \hline
{Nucleus} & 
{Auger $e^-$ } & 
{IC $e^-$} & 
{$e^+$} & 
{X-ray } \\ \hline
        \nuc{57}{Co} &7.594&10.22&0.000&3.598\\
        \nuc{56}{Co} &3.355&0.374&115.7&1.588\\
        \nuc{55}{Fe} &3.973&0.000&0.000&1.635\\
	\nuc{44}{Ti} &3.519&7.064&0.000&0.768\\
	\nuc{44}{Sc} &0.163&0.074&595.8&0.030\\ \hline
\end{tabular}
\end{table}

\subsection{SNe Ia}
\label{subsec:snia}
To date, the thermonuclear SNe with data coverage beyond $500$ days are SNe~1991bg 
\citep{Turatto96}, 1992A and 1999by \citep{cappellaro97}, 2000E \citep{lair06}, 
2000cx \citep{sollerman04} and 2003hv \citep{leloudas2009}.\footnote{For 
two other SNe~Ia with observations extending to $\sim$\,$2000$ days, SN~1991T 
\citep{schmidt94} and SN~1998bu \citep{cappellaro01}, light echoes dominate the 
late-time emission after $\sim$\,$500$ days.} Only three -- SNe~2000cx, 
2001el \citep{stritzinger07} and 2003hv -- have been observed in the near-IR at very late epochs.

SNe~Ia lack an extended envelope, and thus become 
transparent to $\gamma$-rays relatively early. In fact, they enter the 
positron-dominated phase about $150$--$300$ days after the explosion 
\citep{leibundgut92,milne01,sollerman04}. 
Between $300$ and $600$ days, their UVOIR light curves seem to follow 
the \nuc{56}{Co} decay \citep[e.g.][]{stritzinger07}, indicating that the 
majority of the positrons are trapped. 
During this phase, flux is progressively 
shifted from the optical to the near-IR bands, the $JHK$ light curves being 
flat or declining very slowly. Starting at $\sim$\,$600$ days there
are some indications of a slow-down in the $V$ band below the rate expected for 
\nuc{56}{Co} decay with full positron trapping \citep{sollerman04,lair06,leloudas2009}. 
It has been suggested that a heating source other than \nuc{56}{Co}, 
possibly other radioactive species or electron recombination, 
may be needed to explain this slow-down. 
We point out that a slow-down in the light curve is expected when the leptonic energy
injection from the decay of \nuc{57}{Co} becomes a significant contribution.

\begin{figure}
\epsfig{file=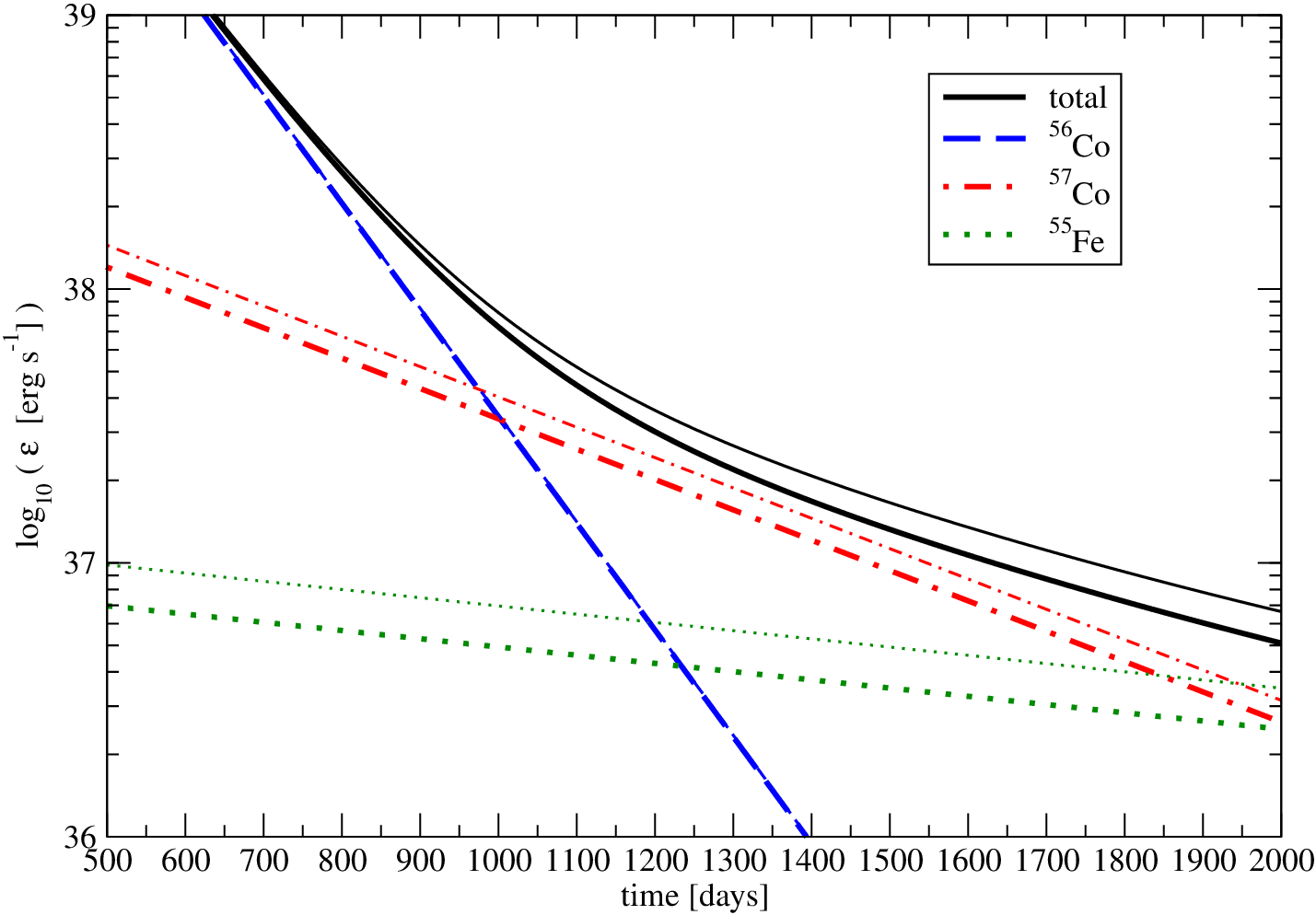,width=8cm}
\caption{\label{fig:1} Instantanous energy generation rates for
  initial abundances taken from the W7 model of \citet{iwamoto99}.
Thick lines are due to positrons, IC and Auger electrons alone. The
thin lines also include the full X-ray dose, which is small
even in the limit of complete X-ray trapping.
The contribution of electrons and positrons from the decays of \nuc{44}{Ti} and \nuc{44}{Sc} is too small to be seen in this figure.}
\end{figure}

To illustrate this, we plot the leptonic energy generation rates of
important long lived isotopes between 500 and 2000 days for
theoretical predictions of the yields of the fiducial W7-model 
\citep{iwamoto99}
for a Type Ia supernova (see Fig.~\ref{fig:1}). 
For this particular choice of yields, the slope of the light curve begins to deviate appreciably from pure
\nuc{56}{Co} decay after about 800 days.
After about 1000 days, the heating due to the electrons produced in
the decay of \nuc{57}{Co} equals the heating from the decay of
\nuc{56}{Co} (mainly due to the positrons). 
It is also noteworthy that the decay of \nuc{55}{Fe} contributes
significantly after about 1400 days (for this model),
but remains below the contribution of \nuc{57}{Co} until about 2000 days. 
The contribution of \nuc{44}{Ti} and its daughter \nuc{44}{Sc}
is below $10^{35}$ erg s$^{-1}$ and remains 
subdominant for many years. 
In reality, the time at which these effects manifest would depend on the 
isotopic ratios synthesized in a particular explosion. Nevertheless, we
predict that a slow-down must eventually occur for reasonable nucleosynthetic yields.

Electron capture radioactive decay also involves the emission of
characteristic fluorescence X-rays (for energies see Table~\ref{tab:2}).
Generally, these X-rays are also neglected as an energy source for the
light curve. 
Since the opacity of the ejecta to X-rays is much higher compared to
$\gamma$-rays, some part of this X-ray radiation may be trapped;
therefore this
contributes to the heating. To properly quantify this would require
detailed radiation transport simulations including realistic X-ray
opacities. However,
even under the extreme assumption that the
X-rays deposit all their energy (see thin lines in Fig.~\ref{fig:1}),
their contribution is small and our conclusions are
not strongly affected.  

\subsection{SN 1998bw}
\label{subsec:1998bw}
SN~1998bw is the first known example of a SN associated with a
$\gamma$-ray burst 
\citep{galama98a}. Spectroscopically it lacked any evidence of H or 
He, and therefore it was classified as a SN Ic. It proved to be an unusually energetic 
explosion of a massive, stripped stellar core ($\geq 10$ \msun; 
\citealt{patat01,nakamura01b,maeda02}). Strong asymmetry was inferred from an 
analysis of nebular spectra \citep{maeda06a}. This asymmetry together with 
the high ejecta velocities and the lack of an extended envelope help to 
reduce the $\gamma$-ray opacity, so that by $1000$ days almost all 
$\gamma$-rays escape freely (the models of \citet{nakamura01a} suggest
that 1998bw enters the positron dominated phase after $\sim400$ days).

Core collapse supernovae synthesize isotopes in different ratios
compared to SNe Ia, in particular they produce more \nuc{44}{Ti}. 
\citet{nakamura01b} calculated nucleosynthetic yields for hypernovae like 1998bw. 
Their 30 Bethe explosion energy, 10 \msun\ He-core model predicts a \nuc{57}{Ni} to 
\nuc{56}{Ni} mass ratio $\mathcal{R}^{57/56} \approx 0.0366$, $\sim
1.5$ times the solar value for the \nuc{57}{Fe} to \nuc{56}{Fe} ratio 
(which is $0.0234$; \citealt{lodders03}).
The isotopic ratios for the asymmetric model of \citet{maeda02} are
similar, albeit with a slightly higher contribution of \nuc{44}{Ti}.  
\citet{hoeflich99} proposed an alternative, highly asymmetric explosion 
scenario that requires only $0.2$ \msun\ of \nuc{56}{Ni}. 

After modeling the UVOIR light curve to $\sim$\,$1000$ days after the 
explosion, \citet{sollerman02b} showed that a simple model without
contributions from freeze-out effects, circumstellar interaction,
accretion by a central compact object or light echoes requires a
 \nuc{57}{Ni} to \nuc{56}{Ni} ratio $\sim 13.5$ times greater than
solar.
This implies a discrepancy of almost one order of magnitude between
the value of $\mathcal{R}^{57/56}$
expected from explosive nucleosynthesis and that suggested by the
light curve modeling. However, as we now demonstrate, this discrepancy
reduces if
IC and Auger electrons are included in the light-curve calculations. 

\begin{figure}
\epsfig{file=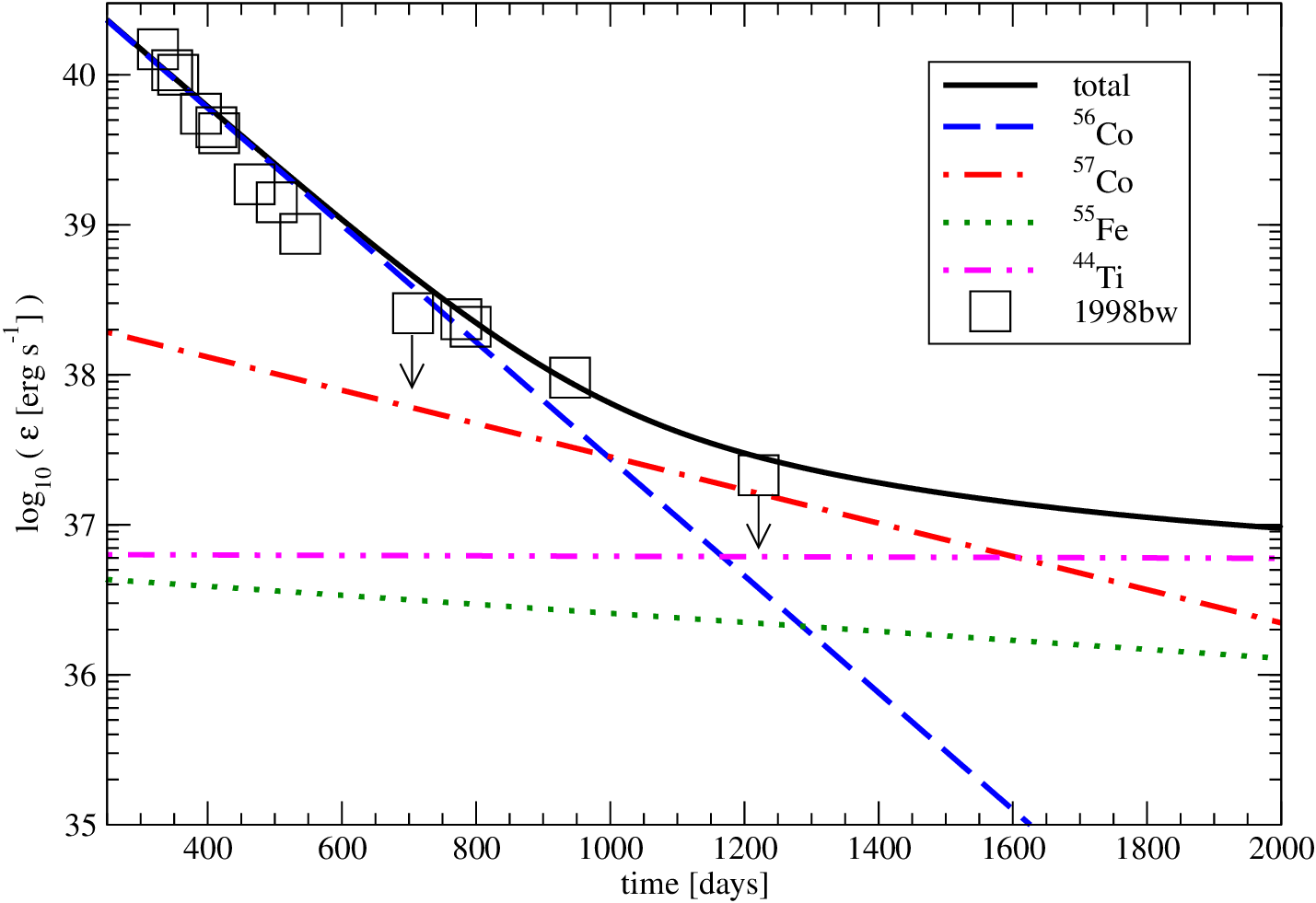,width=8cm}
\caption{\label{fig:2} Instantaneous energy generation rates due to
  radioactivity for initial abundances taken from the 30 Bethe 10
  \msun\ He-core hypernova model of \citet{nakamura01b}. Only leptons
  and X-rays (which are small compared to the leptonic
  contribution) are included (i.e. no $\gamma$-ray contribution or freeze-out
  effects are considered). Squares are the data of 1998bw of
  \citet{sollerman02b}. Arrows represent $3\sigma$ upper limits.
There are no free parameters. }
\end{figure}

In Fig.~\ref{fig:2} we show the combined leptonic and X-ray luminosity
corresponding to yield predictions for the aforementioned \citet{nakamura01b} model.
The non-$\gamma$-ray heating due to \nuc{57}{Co} is the dominant
contribution between $\sim1000$ and 1600 days. 
It is remarkable that this very simplistic approach naturally reproduces the
observed slow-down of the light curve at $\sim900$ days (see
Fig.~\ref{fig:2}) without the need to invoke any abundance
enhancements.

\section{Discussion and conclusions}
\label{sec:discussion}

We conclude that the electrons produced in the decay of \nuc{57}{Co}
constitute a significant energy source for both stripped core collapse
and thermonuclear supernovae
and should be included in codes that model their late-time light
curves.
Fortunately, the vast majority of the energy injected via these
processes is in the form of primary IC/Auger electrons with energy
$\simgt 1$keV. Since the energy deposition channels from such high
energy electrons are independent of the input electron spectrum
\citep{kozma92}, this means that standard methods for treating e.g. 
primary Compton electrons are directly applicable.

We suggest that owing to the neglect of 
the heating due to electrons described in
this paper,
masses previously derived for \nuc{44}{Ti} and \nuc{57}{Ni} from light curves
should be regarded as upper limits.
In particular, a reasonable production factor $\mathcal{R}^{57/56}$ 
naturally explains the slow-down in the lightcurve of 1998bw 
without the need to invoke extreme super solar values.
In addition, we predict that after $\sim1400$
days, the decay of \nuc{55}{Fe} causes a further slow-down of SNe~Ia light
curves, even though it proceeds via a ground-state to ground-state transition.
In a scenario with a large escape fraction of the more energetic
positrons, the effect may be noticeable even earlier.

We limit our discussion in Section~3 to particular classes of SNe in
which the ejecta are expected to be nearly optically thin to
$\gamma$-rays at the relevant epochs ($\sim 1000$~days) so that 
leptonic heating rates are likely to dominate. However, 
\nuc{57}{Co} electrons may also be relevant to more
complex cases, including SNe~II. 
SN~1987A (Type II-P) is the only SN with a bolometric light curve extending from the 
earliest stages all the way to the remnant phase \citep{suntzeff92,suntzeff97}. 
The isotopic ratios of the important radionuclides for SN~1987A are not
very different from the predictions for a hypernova \citep{nakamura01b} and the differential effect of
neglecting the leptonic energy deposition from \nuc{57}{Ni} must act
in the same sense (i.e. it will lead to an
overestimation of $\mathcal{R}^{57/56}$). 
However, this effect is
complicated by the large
envelope mass, which means the ejecta are not fully transparent to 
$\gamma$-rays at the relevant epoch 
\citep{woosley89} and thus leptonic contributions play a smaller role
in heating the ejecta. 
To quantify this requires 
detailed calculations involving both $\gamma$-ray transport and
freeze-out effects \citep{fransson93}, which we defer to future work.

The additional heating channel
pointed out in this paper allows, in principle, for a larger escape fraction
of positrons than previously suggested by modeling late-time light
curves -- as noted in Section~1. Past analyses of light curves have
deduced that the escape fraction must be very small, but any additional
heating source relaxes this constraint somewhat. This may have
implications for understanding the origin of the Galactic population
of positrons and the strength of the Galactic 511 keV annihilation line.

Our calculations
have used atomic data for neutral atoms. 
Although matter in supernova remnants is ionized, for light or
moderate ionization states, we do not expect the
IC and Auger decay energies to change significantly, since 
most of the physics concerns the K, L, and inner M shell. 
For very high levels of ionization, the numbers listed in
Table~\ref{tab:2} are inappropriate. 
However, 
high levels of ionization are not
expected at late times \citep[e.g.][]{kozma98} unless 
the ejecta is shock-ionized by interaction with
circumstellar material, a scenario which would, in any case, likely prohibit
reliable abundance determination because of the need to separate the
contributions from the shock heating and
radioactive decays to the bolometric light curve.

The case of SN~1987A (not a very luminous object, but at a 
distance of only $\sim50$ kpc in the LMC) demonstrates that SN explosions of any type
exploding within the Local Group can be followed for many years after the explosion. 
Adopting an average rate of $\sim$\,0.9 
SNe (100\,yr)$^{-1}$ $(10^{10}\ L^B_\odot)^{-1}$ in spiral galaxies \citep{cappellaro99}, and a 
Local-Group $B$-band luminosity of $\sim$\,$6 \times 10^{10}\ L^B_\odot$ 
\citep{devaucouleurs76,sandage81}, statistically a few Local-Group 
SNe are expected every $100$ years. 
It is therefore plausible that the effects discussed in this paper
can be observed unambiguously in the future. 
 
\section*{Acknowledgments}
We thank the anonymous referee, Jesper Sollerman, Daniel Sauer and Wolfgang Hillebrandt for
reading of the manuscript and their comments. 
We thank Friedrich R\"{o}pke and
 Keiichi Maeda for their helpful discussions. This work is supported
 by the Max-Planck-Institute for Astrophysics, 
the Emmy Noether Program of the German Research Foundation (DFG; RO~3676/1-1)
and the Transregional Collaborative Research Centre TRR 33 `The Dark Universe' of the DFG. 

\bibliographystyle{mn2e}
\bibliography{bibliography}

\end{document}